\documentclass[runningheads]{llncs}
\usepackage[english]{babel}
\usepackage[T1]{fontenc}
\usepackage{amsmath}
\usepackage{mathtools}
\usepackage{amsfonts}
\usepackage{paralist}
\usepackage{enumitem}

\usepackage{hyperref}

\setlist[itemize]{noitemsep, topsep=0pt}
\setlist[enumerate]{noitemsep, topsep=0pt}

\begin{document}
\title{Benchmarking Optimization Solvers and Symmetry Breakers for the Automated Deployment of Component-based Applications\\ in the Cloud\\
\textsc{EXTENDED ABSTRACT}}
\titlerunning{Benchmarking Optimization Solvers and Symmetry Breakers}
\author{Bogdan David \and M\u{a}d\u{a}lina Era\c{s}cu
}
\authorrunning{David et al.}
\institute{Faculty of Mathematics and Informatics\\West University of Timi\c{s}oara\\
\email{\{firstname.lastname\}@e-uvt.ro}}
\maketitle              % typeset the header of the contribution
\begin{abstract}
Optimization solvers based on methods from constraint programming (OR-Tools, Chuffed, Gecode), optimization modulo theory (Z3), and mathematical programming (CPLEX) are successfully applied nowadays to solve many non-trivial examples. However, for solving the problem of \emph{automated deployment} in the Cloud of \emph{component-based applications}, their computational requirements are huge making automatic optimization practically impossible with the current \emph{general} optimization techniques. To overcome the difficulty, we exploited the sweet spots of the underlying problem in order to identify search space reduction methods. We came up with $15$ symmetry breaking strategies which we tested in a static symmetry breaking setting on the solvers enumerated above and on $4$ classes of problems.

As a result, all symmetry breaking strategies led to significant improvement of the computational time of all solvers, most notably, Z3 performed the best compared to the others. 
As an observation, the symmetry breaking strategies confirmed that, when applied in a static setting, they may interact badly with the underlying techniques implemented by the solvers.

\end{abstract}

%
%%%%%%%%%%%%%%%%%%%%%%%%%%%%%%%%%%
%%%%%%%%%% sec:introduction %%%%%%
%%%%%%%%%%%%%%%%%%%%%%%%%%%%%%%%%%
\section{Introduction}\label{sec:intro}
The problem of \emph{automated deployment} in the Cloud of component-based applications received attention due to increased demand of digitalization of businesses. 
It consists of the following steps:
\begin{inparaenum}[\itshape (1)\upshape] 
\item selection of the computing resources, 
\item the distribution/assignment of the application components over the available computing resources, and 
\item its dynamic modification to cope with peaks of user requests. 
\end{inparaenum} In paper \cite{DBLP:journals/jlap/ErascuMZ21}, we tackled only the first two steps of the deployment problem. In particular, our approach was used to synthesize the initial static optimal deployment of the application which consists of an assignment of application components to VMs such that the application functional requirements are fulfilled and costs are minimized. 

The contributions of \cite{DBLP:journals/jlap/ErascuMZ21} are:
\begin{inparaenum}[\itshape (i)\upshape]
\item we formalized the Cloud deployment problem by abstracting the particularities of four classes of real-world problems;
\item we proposed a methodology analyzing the particularities of the problem with the aim of identifying search space reduction methods (these are methods exploiting the symmetries of the general Cloud deployment problem, respectively methods utilizing the graph representation of the interaction between the components of each application);
\item we assessed and compared the performance of two types of tools, namely mathematical programming (CPLEX~\cite{2016_CPLEX_usermanual}) and computational logic (the optimization modulo theory solver Z3~\cite{10.1007/978-3-540-78800-3_24});
\item we identified limits in their scalability and applied six search space reduction methods aiming to improve their performance.
\end{inparaenum} 

This paper extends \cite{DBLP:journals/jlap/ErascuMZ21} in the following ways:
\begin{enumerate}
\item a new formalization in the Minizinc \cite{DBLP:conf/cp/NethercoteSBBDT07} constraint modeling language;
\item the performance comparison adds the constraint programming solvers OR-Tools~\cite{ortools}, Gecode \cite{gecode} and  Chuffed~\cite{chuffed} which are available from the Minizinc IDE;
\item the list of the symmetry breakers from \cite{DBLP:journals/jlap/ErascuMZ21} is enriched with composition of all possible combinations of single symmetry breakers. These symmetry breakers are tested on the constraint programming solvers OR-Tools~\cite{ortools}, Gecode \cite{gecode} and  Chuffed~\cite{chuffed}, optimization modulo theory solver Z3 \cite{DBLP:conf/tacas/BjornerPF15}, and mathematical programming solver CPLEX \cite{2016_CPLEX_usermanual}.
\end{enumerate}

The rest of the paper is organized as follows. Section \ref{sec:scene} briefly introduces the problem and the particularities of each formalization. Section \ref{sec:experimentalanalysis} shows the need for strategies to reduce the search space and, to this aim, introduces symmetry breaking techniques as well as the principles to combine them. Conclusions of the experimental analysis are presented in Section \ref{sec:discussion}. 
%%%%%%%%%%%%%%%%%%%%%%%%%%%%%%%%%%
%%%% sec:Setting the scene %%%%%%
%%%%%%%%%%%%%%%%%%%%%%%%%%%%%%%%%%
\section{Setting the Scene }\label{sec:scene}
\paragraph*{Problem Definition} The description of the problem first appeared in \cite{DBLP:journals/jlap/ErascuMZ21}. We have $N$ interacting components, \mbox{$C=\{C_1,\ldots, C_N\}$}, to be assigned to a set of $M$ virtual machines, $V=\{V_1, \ldots, V_M\}$. Each component $C_i$ is characterized by a set of requirements concerning the hardware resources. Each virtual machine, $V_k$, is characterized by a \emph{type}, which is comprised by hardware/software characteristics and leasing price. There are also \emph{structural constraints} describing the interactions between components. The problem is to find: 
\begin{enumerate}
	\item an assignment matrix $a$ with binary entries $a_{ik}\in \{0,1\}$ for $i=\overline{1,N}$, $k=\overline{1,M}$, which are interpreted as follows: 
	$a_{ik}= 1 \hbox{ if }C_{i} \hbox{ is assigned to } V_k, \text{ and } 0$, otherwise; and
	\item the type selection vector $\emph{t}$ with integer entries $\emph{t}_k$ for $k=\overline{1,M}$, representing the type (from a predefined set) of each VM leased;
\end{enumerate}
such that:
\begin{inparaenum}[\itshape (i)\upshape]
	\item the structural constraints, and
	\item the hardware requirements  (capacity constraints) of all components are satisfied; and 
	\item the purchasing/ leasing price is minimized. 
\end{inparaenum}

The \emph{structural constraints} are \emph{application-specific} and derived in accordance with the analysis of the case studies. These are:
\begin{itemize}
    \item \emph{Conflict:} components in conflict cannot be deployed on the same VM. 
    \item \emph{Co-location:} components in co-location must be deployed on the same VM. 
    \item \emph{Exclusive deployment:} Only one of the components in exclusive deployment must be deployed in the same deployment plan.
    \item \emph{Require-Provide:} one component requires or provides some functionalities offered, respectively provides, of another. Such an interaction induces constraints on the number of instances corresponding to the interacting components as follows.
    \item \emph{Full deployment:} components in this relationship must be deployed on all VMs leased, except those which would induce conflicts between components.
    \item \emph{Deployment with bounded number of instances} occur when the number of instances of deployed components must be equal, greater or less than some values.
\end{itemize}

\emph{General constraints} are always considered in the formalization and are related to the: 
\begin{inparaenum}[\itshape (i)\upshape]
	\item \emph{basic allocation} rules, 
	\item \emph{occupancy} criteria,
	\item hardware \emph{capacity} of the VM offers, and
	\item \emph{link} between the VM offers and the components hardware/software requirements.
\end{inparaenum}

We stated the problems as a linear constraint optimization problem (COP). We redirect the reader to \cite{DBLP:journals/jlap/ErascuMZ21} for a full description of it.

\paragraph*{Problem Formalization}
The formalization for all three types of solvers, that is constraint programming solvers, SMT solvers, and mathematical programming solver, has almost a one-to-one correspondence between linear constraints present in the definition of COP and the implementation. We did not apply optimizations exploiting the particularities of the modelling languages because we wanted to have a fair comparison of the different formalisms. However, in the future work we plan to take advantage of their sweet spots.

\paragraph*{Minizinc models} The Minizinc models  are the ones newly introduced in this paper. They are organized as follows:
\begin{inparaenum}[\itshape (i)\upshape]
\item there are surrogate models for each problem in which the maximum number of needed VMs is computed; %can we have one general and the applications instantiate the ones they need?
\item there is a model gathering together all constraints, both general and application specific (the model corresponding to each application instantiates the constraints needed for its modeling)
\item there is a model gathering together all symmetry breakers developed which are then instantiated based on the tests which want to be performed.
\end{inparaenum}

%%%%%%%%%%%%%%%%%%%%%%%%%%%%%%%%%%%%%%%%%%%%%%%%%%%%%%%
%%##%%%%%%%%%% sec: Experimental Analysis %%%%%%%%%%%%%%
%%%%%%%%%%%%%%%%%%%%%%%%%%%%%%%%%%%%%%%%%%%%%%%%%%%%%%%
\section{Experimental Analysis}\label{sec:experimentalanalysis}
The principles of the experimental analysis were introduced in \cite{DBLP:journals/jlap/ErascuMZ21}: on one hand, we want to assess the \emph{scalability} of state-of-the-art general CP (Chuffed~\cite{chuffed}, Gecode~\cite{gecode}, OR-Tools~\cite{ortools}), MP (CPLEX~\cite{2016_CPLEX_usermanual}) and OMT (Z3~\cite{DBLP:conf/tacas/BjornerPF15}) tools in solving COPs corresponding to realistic case studies. On the other hand, we evaluate the \emph{effectiveness} of various static symmetry breaking techniques in improving the computational time of solving these problems (see Section~\ref{sec:ExperimentalSettings}). This is because tests (see Tables~\ref{tab:nosymbreakingWordpress}-\ref{tab:nosymbreakingOryx2}) revealed that the naive application of general CP, MP and OMT techniques is not sufficient to solve realistic Cloud deployment applications.

We consider four case studies (Secure Web Container, Secure Billing Email Service, Oryx2, and Wordpress) which exhibit:
\begin{inparaenum}[\itshape (i)\upshape]
	\item different hardware characteristics of components and the rich interactions type in between (structural constraints);
	\item the kind of linear constraints used to formalize the problem; and
	\item the kind of solution we are searching for.
\end{inparaenum}
A full description of these case studies is in~\cite{DBLP:journals/jlap/ErascuMZ21}.

The scalability and effectiveness are evaluated from two perspectives: number of VM offers, respectively number of deployed instances of components.  For \emph{Secure Web Container}, \emph{Secure Billing Email} and \emph{Oryx2} applications, we considered up to 500 VM offers. Additionally, for the \emph{Wordpress} application, we considered up to $12$ instances of the Wordpress component to be deployed. The set of offers was crawled from the Amazon CPs offers list. 
%%%%%%%%%%%%%%%%%%%%%%%%%%%%%%%%%%%%%%%%%%%%%%%%%%
%%%%%%%%%% sec: Experimental Settings %%%%%%%%%%%%%%
%%%%%%%%%%%%%%%%%%%%%%%%%%%%%%%%%%%%%%%%%%%%%%%%%%
\subsection{Experimental Settings}\label{sec:ExperimentalSettings}
%In this section we present and motivate the plethora of symmetry breakers tested on two types of optimization tools for optimization, as well as the characteristics of the hardware we ran the experiments on.
%%%%%%%%%%%%%%%%%%%%%%%%%
%%%%%% Selected Symmetry Breaking Strategies %%%%%%%%%
%%%%%%%%%%%%%%%%%%%%%%%%%
\subsubsection{Selected Symmetry Breaking Strategies}\label{sec:SelectedSymmetryBreakingStrategies}
Aiming to reduce the search space size, a set of strategies have been selected in order to exploit the particularities of the problem:
\begin{inparaenum}[\itshape (i)\upshape] 
\item the VMs needed for application deployment might have different characteristics; 
\item applications components might be in conflict hence conflict-type constraints can be exploited; 
\item the number of instances to be deployed is unknown.
\end{inparaenum}

Our approach is incremental and experimental: we start with traditional symmetry breakers that have been used for other problems related to bin-packing and combine them with the aim of further search space reduction.
%%%%%%%%%%%%%%%%%%%%%%%%%%%%%%%%%
%%% Simple symmetry breakers %%%%
%%%%%%%%%%%%%%%%%%%%%%%%%%%%%%%%%
\paragraph{Simple symmetry breakers}
\begin{description}
\item{\emph{Price-based ordering} (PR)}. This strategy aims to break symmetry by ordering  the vector containing the types of used VMs decreasingly by price, i.e. 
%the price of the VM corresponding to column $j$ is greater or equal than the price of the VM corresponding to column $(j+1)$.  
%\begin{equation} \label{eq:PriceOrderOnVMs}
$p_k\geq p_{k+1},\   k=\overline{1,M-1}.$ This means that the solution will be characterized by the fact that the columns of the assignment matrix will be ordered decreasingly by the price of the corresponding VMs.
%\end{equation}
\item{\emph{Lexicographic ordering} (LX)}. This corresponds to the traditional strategy aiming to break column-wise symmetries. The constraints to be added aiming to ensure that two columns, $k$ and $(k+1)$ are in a decreasing lexicographic order, i.e. $a_{*k}\succ_{lex} a_{*(k+1)}$, are 
$\bigwedge\limits_{l=1}^{i-1} (a_{lk}=a_{l(k+1)}) \Longrightarrow  a_{ik}\ge a_{i(k+1)}, \ \forall i=\overline{1,N}$.
\item{\emph{Load-based ordering (L)}}. This is a column-wise symmetry breaker which orders decreasingly the columns by the number of the component instances they accommodate: $
\sum_{i=1}^N a_{ik}\geq \sum_{i=1}^N a_{i(k+1)}, \ k=\overline{1,M-1}$.
\item{\emph{Fixed values (FV).}} The search space can be reduced also by fixing the values of some variables starting from the application specific constraints, in particular conflict constrains. More precisely, the graph composed by the components being in conflict is used to identify components which must be placed on different machines and hence the values of the corresponding decision variables are fixed. The identification of these components is done by constructing the clique with maximum deployment size.
\end{description}
%%%%%%%%%%%%%%%%%%%%%%%%%%%%%%%%%
%% Composed symmetry breakers %%%
%%%%%%%%%%%%%%%%%%%%%%%%%%%%%%%%%
\paragraph{Composed symmetry breakers}
The symmetry breakers above can be composed leading to the following symmetry breakers:
\begin{itemize}
\item FV-PR, FV-L, FV-LX, PR-L, PR-LX, L-PR, L-LX,
\item FV-PR-L, FV-PR-LX, FV-L-PR, FV-L-LX, PR-L-LX, L-PR-LX,
\item FV-PR-L-LX, FV-L-PR-LX
\end{itemize}
These symmetry breakers are so the subsequent breaks ties of the former. For example, FV-PR fixes on separate machines the decision variables corresponding to the component instances being in the clique with maximum deployment size and the machines left unoccupied are ordered decreasingly by price. In the case of PR-L-LX, the machines are ordered decreasingly by price, those with the same price are ordered decreasingly by the number of components they host and those with the same number of instances are ordered lexicographically. 

It is worth noticing that the symmetry breakers involving FV must apply FV the very first. This is because FV is used as a preprocessing step which has a positive impact on the solvers as it introduces equalities. %We did not experiment with symmetry breakers applying FV on the later positions as we believe those compositions are not natural and do not have a computational impact.
\subsubsection{Software and Hardware Settings}\label{sec:SW&HDD-design}
We used Minizinc v0.7.0 as the constraint modeling language. We mention that the Minizinc models follow the formalization and no optimizations were performed because we wanted to be as close as possible to the OMT and CPLEX formalizations in order to have a fair computational comparison between the newly considered solvers and the existing results. The CP solvers used (Chuffed, Gecode, OR-Tools) are called from Minizinc IDE with the default values for parameters. The OMT formalization is done using the Z3 Python API and uses quantifier-free linear integer arithmetic. Z3 was used with the default values of the parameters. In the case of the mathematical programming solver CPLEX, we used the Python API with the no symmetry breaking option manually activated.

The source code and the experimental results are available online at \url{https://github.com/BogdanD02/Cloud-Resource-Provisioning}, release v1.0.0. All reported timings are in seconds. They only include the actual solving time of the optimization problem and not the pre-processing steps.

All tests in this paper were performed on an  Intel(R) Core (TM) i5-9400F CPU @ 3.90GHz using Chuffed v0.10.4, Gecode v6.3.0, OR-Tools v9.0.0, CPLEX v12.9.0 and Z3 v4.10.2. 

%%%%%%%%%%%%%%%%%%%%%%%%%%%%%%%%%%%%%%%%%%%%%%%%%%
%%%%%%%%%%%%%%%%%%%% sec: Results %%%%%%%%%%%%%%%
%%%%%%%%%%%%%%%%%%%%%%%%%%%%%%%%%%%%%%%%%%%%%%%%%%
\subsection{Results}\label{sec:results}

Tables~\ref{tab:nosymbreakingWordpress}-\ref{tab:nosymbreakingSecureWeb} include the results obtained without using symmetry breaking strategies. The list of offers (columns \#o) was crawled from the Amazon site\footnote{\url{https://aws.amazon.com/}}. Each list of VM offers covers the main instance types, for example, \texttt{small}, \texttt{medium}, \texttt{large}. The list of offers can be viewed as a containment hierarchy (i.e. the list of 20 offers is included in the list of 40 offers etc.).
\begin{table}[t]
\begin{center}
\caption{Scalability tests for Wordpress with different instances (\#i). Time values are expressed in seconds.}
\label{tab:nosymbreakingWordpress}
\begin{tabular}{|ccccc||cccc}
\hline
\multicolumn{1}{|c|}{\textbf{\#i}} & \multicolumn{1}{c|}{\textbf{\#o=20}} & \multicolumn{1}{c|}{\textbf{\#o=40}} & \multicolumn{1}{c|}{\textbf{\#o=250}} & \textbf{\#o=500} & \multicolumn{1}{c|}{\textbf{\#o=20}} & \multicolumn{1}{c|}{\textbf{\#o=40}} & \multicolumn{1}{c|}{\textbf{\#o=250}} & \multicolumn{1}{c|}{\textbf{\#o=500}} \\ \hline
\multicolumn{5}{|c|}{\textbf{OR-Tools}}                                                                                                                                    & \multicolumn{4}{c|}{\textbf{CPLEX}}                                                                                                                         \\ \hline
\multicolumn{1}{|c|}{\textbf{3}}  & \multicolumn{1}{c|}{3.49}            & \multicolumn{1}{c|}{8.38}            & \multicolumn{1}{c|}{96.05}            & 191.04           & \multicolumn{1}{c|}{9.66}            & \multicolumn{1}{c|}{-}               & \multicolumn{1}{c|}{-}                & \multicolumn{1}{c|}{-}                \\ \hline
\multicolumn{1}{|c|}{\textbf{4}}  & \multicolumn{1}{c|}{23.25}           & \multicolumn{1}{c|}{56.07}           & \multicolumn{1}{c|}{501.43}           & 987.91           & \multicolumn{1}{c|}{121.96}          & \multicolumn{1}{c|}{-}               & \multicolumn{1}{c|}{-}                & \multicolumn{1}{c|}{-}                \\ \hline
\multicolumn{1}{|c|}{\textbf{5}}  & \multicolumn{1}{c|}{149.47}          & \multicolumn{1}{c|}{425.03}          & \multicolumn{1}{c|}{-}                & -                & \multicolumn{1}{c|}{446.02}          & \multicolumn{1}{c|}{-}               & \multicolumn{1}{c|}{-}                & \multicolumn{1}{c|}{-}                \\ \hline
\multicolumn{1}{|c|}{\textbf{6}}  & \multicolumn{1}{c|}{493.39}          & \multicolumn{1}{c|}{1173.46}         & \multicolumn{1}{c|}{-}                & -                & \multicolumn{1}{c|}{664.68}          & \multicolumn{1}{c|}{-}               & \multicolumn{1}{c|}{-}                & \multicolumn{1}{c|}{-}                \\ \hline
\multicolumn{5}{|c|}{\textbf{Gecode}}                                                                                                                                      & \multicolumn{4}{c|}{\textbf{Chuffed}}                                                                                                                       \\ \hline
\multicolumn{1}{|c|}{\textbf{3}}  & \multicolumn{1}{c|}{2.13}            & \multicolumn{1}{c|}{2.19}            & \multicolumn{1}{c|}{-}                & -                & \multicolumn{1}{c|}{2.05}            & \multicolumn{1}{c|}{3.7}             & \multicolumn{1}{c|}{45.88}            & \multicolumn{1}{c|}{447.56}           \\ \hline
\multicolumn{1}{|c|}{\textbf{4}}  & \multicolumn{1}{c|}{14.84}           & \multicolumn{1}{c|}{23.83}           & \multicolumn{1}{c|}{-}                & -                & \multicolumn{1}{c|}{23.73}           & \multicolumn{1}{c|}{114.18}          & \multicolumn{1}{c|}{1866.61}          & \multicolumn{1}{c|}{-}                \\ \hline
\multicolumn{1}{|c|}{\textbf{5}}  & \multicolumn{1}{c|}{162.13}          & \multicolumn{1}{c|}{-}               & \multicolumn{1}{c|}{-}                & -                & \multicolumn{1}{c|}{531.19}          & \multicolumn{1}{c|}{2278.76}         & \multicolumn{1}{c|}{-}                & \multicolumn{1}{c|}{-}                \\ \hline
\multicolumn{5}{|c|}{\textbf{Z3}}                                                                                                                                          &                                      &                                      &                                       &                                       \\ \cline{1-5}
\multicolumn{1}{|c|}{\textbf{3}}  & \multicolumn{1}{c|}{2.82}            & \multicolumn{1}{c|}{4.13}            & \multicolumn{1}{c|}{103.19}           & 391.87           &                                      &                                      &                                       &                                       \\ \cline{1-5}
\multicolumn{1}{|c|}{\textbf{4}}  & \multicolumn{1}{c|}{46.46}           & \multicolumn{1}{c|}{275.81}          & \multicolumn{1}{c|}{-}                & -                &                                      &                                      &                                       &                                       \\ \cline{1-5}
\end{tabular}
\end{center}
\end{table}
%%%%%%%%%%%%%%%%%%%%%%%%%%%%%%%%%%%%%%%%%%%%%%%%
\begin{table}[t]
\begin{center}
\caption{Scalability tests for Oryx2. Time values are expressed in seconds.}
\label{tab:nosymbreakingOryx2}
\begin{tabular}{|cccc||cccc}
\hline
\multicolumn{1}{|c|}{\textbf{\#o=20}} & \multicolumn{1}{c|}{\textbf{\#o=40}} & \multicolumn{1}{c|}{\textbf{\#o=250}} & \textbf{\#o=500} & \multicolumn{1}{c|}{\textbf{\#o=20}} & \multicolumn{1}{c|}{\textbf{\#o=40}} & \multicolumn{1}{c|}{\textbf{\#o=250}} & \multicolumn{1}{c|}{\textbf{\#o=500}} \\ \hline
\multicolumn{4}{|c|}{\textbf{OR-Tools}}                                                                                                 & \multicolumn{4}{c|}{\textbf{CPLEX}}                                                                                                                         \\ \hline
\multicolumn{1}{|c|}{0.95}            & \multicolumn{1}{c|}{1.12}            & \multicolumn{1}{c|}{4.25}             & 5.79             & \multicolumn{1}{c|}{0.16}            & \multicolumn{1}{c|}{0.54}            & \multicolumn{1}{c|}{-}                & \multicolumn{1}{c|}{-}                \\ \hline
\multicolumn{4}{|c|}{\textbf{Gecode}}                                                                                                   & \multicolumn{4}{c|}{\textbf{Chuffed}}                                                                                                                       \\ \hline
\multicolumn{1}{|c|}{70.99}           & \multicolumn{1}{c|}{104.54}          & \multicolumn{1}{c|}{234.27}           & 465.72           & \multicolumn{1}{c|}{128.67}          & \multicolumn{1}{c|}{154.11}          & \multicolumn{1}{c|}{294.25}           & \multicolumn{1}{c|}{396.58}           \\ \hline
\multicolumn{4}{|c|}{\textbf{Z3}}                                                                                                       &                                      &                                      &                                       &                                       \\ \cline{1-4}
\multicolumn{1}{|c|}{13.35}           & \multicolumn{1}{c|}{15.36}           & \multicolumn{1}{c|}{453.2}            & 717.99           &                                      &                                      &                                       &                                       \\ \cline{1-4}
\end{tabular}
\end{center}
\end{table}
%%%%%%%%%%%%%%%%%%%%%%%%%%%%%
\begin{table}[t]
\begin{center}
\caption{Scalability tests for Secure Billing Email Service. Time values are expressed in seconds.}
\label{tab:nosymbreakingSecureBilling}
\begin{tabular}{|cccc||cccc}
\hline
\multicolumn{1}{|c|}{\textbf{\#o=20}} & \multicolumn{1}{c|}{\textbf{\#o=40}} & \multicolumn{1}{c|}{\textbf{\#o=250}} & \textbf{\#o=500} & \multicolumn{1}{c|}{\textbf{\#o=20}} & \multicolumn{1}{c|}{\textbf{\#o=40}} & \multicolumn{1}{c|}{\textbf{\#o=250}} & \multicolumn{1}{c|}{\textbf{\#o=500}} \\ \hline
\multicolumn{4}{|c|}{\textbf{OR-Tools}}                                                                                                 & \multicolumn{4}{c|}{\textbf{CPLEX}}                                                                                                                         \\ \hline
\multicolumn{1}{|c|}{0.85}            & \multicolumn{1}{c|}{1.22}            & \multicolumn{1}{c|}{8.04}             & 18.68            & \multicolumn{1}{c|}{0.73}            & \multicolumn{1}{c|}{4.73}            & \multicolumn{1}{c|}{378.47}           & \multicolumn{1}{c|}{-}                \\ \hline
\multicolumn{4}{|c|}{\textbf{Gecode}}                                                                                                   & \multicolumn{4}{c|}{\textbf{Chuffed}}                                                                                                                       \\ \hline
\multicolumn{1}{|c|}{0.73}            & \multicolumn{1}{c|}{1.42}            & \multicolumn{1}{c|}{76.63}            & 119.32           & \multicolumn{1}{c|}{0.79}            & \multicolumn{1}{c|}{0.82}            & \multicolumn{1}{c|}{1.35}             & \multicolumn{1}{c|}{3.11}             \\ \hline
\multicolumn{4}{|c|}{\textbf{Z3}}                                                                                                       &                                      &                                      &                                       &                                       \\ \cline{1-4}
\multicolumn{1}{|c|}{0.34}            & \multicolumn{1}{c|}{0.70}            & \multicolumn{1}{c|}{3.37}             & 8.29             &                                      &                                      &                                       &                                       \\ \cline{1-4}
\end{tabular}
\end{center}
\end{table}
%%%%%%%%%%%%%%%%%%%%%%%%%%%%%
\begin{table}[t]
\begin{center}
\caption{Scalability tests for Secure Web Container. Time values are expressed in seconds.}
\label{tab:nosymbreakingSecureWeb}
\begin{tabular}{|cccc||cccc}
\hline
\multicolumn{1}{|c|}{\textbf{\#o=20}} & \multicolumn{1}{c|}{\textbf{\#o=40}} & \multicolumn{1}{c|}{\textbf{\#o=250}} & \textbf{\#o=500} & \multicolumn{1}{c|}{\textbf{\#o=20}} & \multicolumn{1}{c|}{\textbf{\#o=40}} & \multicolumn{1}{c|}{\textbf{\#o=250}} & \multicolumn{1}{c|}{\textbf{\#o=500}} \\ \hline
\multicolumn{4}{|c|}{\textbf{OR-Tools}}                                                                                                 & \multicolumn{4}{c|}{\textbf{CPLEX}}                                                                                                                         \\ \hline
\multicolumn{1}{|c|}{1.50}            & \multicolumn{1}{c|}{4.85}            & \multicolumn{1}{c|}{28.79}            & 57.33            & \multicolumn{1}{c|}{2.68}            & \multicolumn{1}{c|}{49.64}           & \multicolumn{1}{c|}{-}                & \multicolumn{1}{c|}{-}                \\ \hline
\multicolumn{4}{|c|}{\textbf{Gecode}}                                                                                                   & \multicolumn{4}{c|}{\textbf{Chuffed}}                                                                                                                       \\ \hline
\multicolumn{1}{|c|}{10.15}           & \multicolumn{1}{c|}{43.84}           & \multicolumn{1}{c|}{76.63}            & 779.44           & \multicolumn{1}{c|}{0.89}            & \multicolumn{1}{c|}{1.01}            & \multicolumn{1}{c|}{2.64}             & \multicolumn{1}{c|}{4.54}             \\ \hline
\multicolumn{4}{|c|}{\textbf{Z3}}                                                                                                       &                                      &                                      &                                       &                                       \\ \cline{1-4}
\multicolumn{1}{|c|}{0.45}            & \multicolumn{1}{c|}{1.14}            & \multicolumn{1}{c|}{8.28}             & 17.48            &                                      &                                      &                                       &                                       \\ \cline{1-4}
\end{tabular}
\end{center}
\end{table}
The tables include only those cases for which we obtained a result in a 40 minutes timeframe. The missing values ({\tiny $-$}) mean that no solution is returned in this timeframe. 

One can observe that CPLEX scales the worst for all case studies while OR-Tools the best. However, none of the tools scale for Wordpress with more than 6 instances and several dozens of offers. 

To overcome the lack of scalability issue, we applied the symmetry breaking strategies described in Section \ref{sec:SelectedSymmetryBreakingStrategies}.
%%%%%%%%%%%%%%%%%%%%%%%%%%%%%%%%%%%%%%
%%%%%%%%% sec: Discussion of the Results %%%%%%%%%%%%
%%%%%%%%%%%%%%%%%%%%%%%%%%%%%%%%%%%%%%
\section{Discussion of the Results}\label{sec:discussion} We conducted various tests involving $5$ solvers and $15$ symmetry breakers. Because of lack of space, we can not include all of them here and we direct the reader to check them in the folder \texttt{PlotData} from release \texttt{v1.0.0} at \url{https://github.com/BogdanD02/Cloud-Resource-Provisioning}.

We draw the following remarks:
\begin{enumerate}
\item Using the $15$ symmetry breakers, out of the $5$ solvers, the best one from the computational time point of view is Z3.
\item For the virtual best solver, that is Z3, the best symmetry breaker is FVPR. This is because we have many Wordpress files to be analyzed ($40$ files corresponding to Wordpress with $3$ up to $12$ instances and $20$, $40$, $250$, $500$ offers), compared to the other applications ($4$ files corresponding to $20$, $40$, $250$, $500$ offers for each of the other applications) for which reduction methods which exploit the graph representation, that is FV, is of benefit. 
\item We also considered the best symmetry breaker for Z3 for each of the case studies. In case of Secure Web Container and Secure Billing Email Service applications the best is FVL, while for Oryx2 is FVLX. There is no surprise that symmetry breakers involving FV give best results, however for more reliable results we should consider more test cases for Secure Billing, Secure Web and Oryx2, since now there are only 4 files analyzed corresponding to different number of offers. We plan to run more tests for each of the case studies for a more accurate analysis.
\item One would expect that the best symmetry breaker is one composing a higher number of individual symmetry breakers as more symmetries are broken so the search space is significantly reduced. However, this is not true: FVPR, composing $2$ symmetry breakers, is better than those composing $3$ or $4$. An explanation for this is, on one hand the number of added constraints which influence the solving time, on the other hand, when using static symmetry breaking, the symmetry breakers can interact badly with the SMT solvers which we used as black box.
\end{enumerate}

\bibliographystyle{splncs04}
\bibliography{mybib}
\end{document}